\documentclass[journal]{IEEEtran}

\ifCLASSINFOpdf
\else
    \usepackage[dvips]{graphicx}
\fi
\usepackage{url}
\usepackage[utf8]{inputenc}
\usepackage{amsfonts}
\usepackage{amsmath,amssymb,epsfig}
\usepackage{enumerate,amsthm,epstopdf}
\usepackage{dsfont}
\usepackage{algpseudocode}
\usepackage{algorithmicx}
\usepackage{algorithm}
\usepackage{listings} 
\usepackage{bm} 
\usepackage{color}
\usepackage{hyperref}
\hypersetup{colorlinks,linkcolor={blue},citecolor={red},urlcolor={blue}}

\usepackage[makeroom]{cancel}
 \usepackage{authblk}
\usepackage{verbatim}

\usepackage{multirow}
\usepackage{rotating}

\usepackage{lineno}
\usepackage{siunitx}
\usepackage{etoolbox}
\newcommand{\ubold}{\fontseries{b}\selectfont}
\robustify\ubold

\DeclareMathOperator{\prox}{prox}

\newcommand{\argmin}{\operatornamewithlimits{argmin}}

\def\bA{{\mathbf A}}
\def\bL{{\mathbf L}}
\def\x{{\mathbf x}}
\def\y{{\mathbf y}}
\def\z{{\mathbf z}}

\def\sX{{\mathsf X}}

\def\bR{{\mathbb R}}

\def\f0{{\mathbf 0}}

\newcommand{\bSigma}{\bm{\Sigma}}

\newcommand{\bmu}{\bm{\mu}}
\newcommand{\bnu}{\bm \nu}

\newcommand{\bxi}{\bm{\xi}}
\newcommand{\bzeta}{\bm{\zeta}}

\theoremstyle{definition}

\newcommand{\cblue}{\textcolor{black}}

\newcommand{\Real}{\mathbb{R}}

\newcommand\acro{PNAIS}

\begin{document}

\title{A Proximal Newton Adaptive Importance Sampler}

\author{V\'ictor Elvira, \IEEEmembership{Senior, IEEE}, \'Emilie Chouzenoux \IEEEmembership{Senior, IEEE}, O. Deniz Akyildiz 
\thanks{E.C. is with the team-project OPIS, Inria Saclay, University Paris-Saclay. E.C. acknowledges support from the European Research Council Starting
Grant MAJORIS ERC-2019-STG-85092. V.~E. is with School of Mathematics, University of Edinburgh. The work of V. E. is supported by ARL/ARO under grant W911NF-22-1-0235. O.~D.~A is with Department of Mathematics, Imperial College London.
} 
}

\maketitle

\begin{abstract}
Adaptive importance sampling (AIS) algorithms are a rising methodology in signal processing, statistics, and machine learning. An effective adaptation of the proposals is key for the success of AIS. Recent works have shown that gradient information about the involved target density can greatly boost performance, but its applicability is restricted to differentiable targets. In this paper, \cblue{we propose a proximal Newton adaptive importance sampler for the estimation of expectations with respect to non-smooth target distributions}. We implement a scaled Newton proximal gradient method to adapt the proposal distributions, enabling efficient and optimized moves even when the target distribution lacks differentiability. \cblue{We show the good performance of the algorithm in two scenarios: one with convex constraints and another with non-smooth sparse priors.}
\end{abstract}


\IEEEpeerreviewmaketitle

\section{Introduction}
Statistical signal processing applications often require Monte Carlo methods to approximate distributions and integrals, e.g., inference in state-space models \cite{djuric2003particle}, rare event estimation \cite{liao2024rare}, or in more general Bayesian inference tasks \cite{candy2016bayesian}. Importance sampling (IS) is a well-known technique to tackle such problems \cite{elvira2021advances}, particularly when direct Monte Carlo from the target distribution is unfeasible (e.g. in Bayesian inference) or when it is inefficient. The performance of IS depends highly on the proposal quality \cite{akyildiz2021convergence,akyildiz2024global}, which has to be chosen appropriately w.r.t. the target density. One popular way to automatize IS is to adapt one or several proposals over an iterative process. In particular, this class of algorithms is called \textit{adaptive importance sampling} (AIS), e.g., see \cite{bugallo2017adaptive} for a review. AIS methods are importance samplers with time-changing (improving) proposal distributions. More precisely, AIS methods specify a sequence of proposal densities, which are adapted over time, and perform importance sampling to estimate expectations with respect to a targeted density. 
\cblue{These methods are widely used in Bayesian inference, where the target density is the posterior distribution, making AIS particularly suited for Bayesian signal processing and machine learning \cite{luengo2020survey,kviman2022multiple}.}

The biggest challenge of AIS methods is to design proposals which move probability mass to the regions of interest under the target density. For certain scenarios, proposing samples which have high probability under target is of certain interest. Since proposal design includes a certain flexibility, using information from the target density, has become a popular route to take. In particular, one can use gradient or Hessian information \cite{elvira2015gradient,schuster2015gradient,elvira2019langevin,elvira2023gradient}. \cblue{Stochastic gradient approximations have also been used, computed with weighted samples \cite{el2019variational,el2019stochastic} or via reinforcement learning~\cite{el2021policy}.}
%
%
When applied to Bayesian inference, a main difficulty arises, however, in models where the log-likelihood or the log-prior is nonsmooth (not differentiable), e.g., in models where the prior promotes sparsity, e.g., in sparse linear regression \cite{mateos2010distributed} or in other Bayesian tasks  with non-differentiable priors (see for instance \cite{goldman2022gradient}). In particular, many interesting test functions, such as indicator functions of sets, are not differentiable. \cblue{In these cases, gradient adaptive methods are not effectively applicable. Variable transformation or projection approaches have been used to deal with simple constraints. The former however often yields an artificial deformation of the initial target, often detrimental to sampler efficiency. The latter, initially dedicated to constrained problems, can be naturally extended to any convex non-smooth function, by using its Moreau-Yosida's envelope, through a proximal step~\cite{combettes2011proximal}. \cblue{In \cite{shukla2025mcmc}, a framework also leveraging Moreau-Yosida envelopes to approximate non-differentiable target distributions has been recently proposed. This proximal approach enables the use of gradient-based MCMC methods in conjunction with importance sampling for improved efficiency of the estimators.} Subgradient updates could also be employed, but these are often suboptimal in terms of convergence speed, difficult to compute for complex targets, making proximal step a more suitable and versatile approach (see discussions in \cite{Habring,DurmusJMLR, encinar2024proximal}, in the context of Langevin MCMC)}. 
   
In this paper, we propose a class of adaptive importance samplers, called \textit{proximal Newton adaptive importance sampler} (PNAIS), which can use available  information of the target efficiently. \cblue{PNAIS can handle a large class of} non-smooth targets, and still efficiently use first and second order information of some terms in the target to propose samples. Our algorithm can be thought of as the importance sampling counterpart of the proximal MCMC method proposed in \cite{corbineau:hal-02474585} (itself improving upon \cite{pereyra2016proximal}), where we integrate several safe rules to cope with non log-concave targets.  
%
{
Without loss of generality, we limit ourselves here to differentiable likelihoods and nonsmooth priors, although the reverse construction would be straightforward to handle.} We discuss the algorithmic choices based on theoretical justifications. We provide a numerical analysis in two challenging examples where subset and sparsity constraints turn off-the-shelf methods unfeasible.

The rest of the paper is organized as follows. In Section \ref{sec_bg}, we provide background about convex optimization and adaptive importance sampling. In Section \ref{sec_contribution}, we described the proposed algorithm. In Section \ref{sec_numerical} we provide numerical results, and we conclude in Section \ref{sec_conclusion}.

\section{Background}
\label{sec_bg}

\subsection{Problem statement}

We aim at estimating the integrals of the following form: 
\begin{align}\label{eq:ProbDef}
\int_\sX \varphi(\x) \widetilde\pi(\x) \mbox{d}\x,
\end{align}
where $\widetilde\pi(\x)$ is the target density defined on  $\sX\subseteq\bR^{d_x}$ and $\varphi(\x)$ is an integrable function.
We focus on targets of the form
\begin{align}\label{eq:targetConvex}
\widetilde\pi(\x) \propto \exp(- f(\x) - g(\x)) \equiv \pi(\x),
\end{align}
where $f$ is differentiable, and $g$ is  not differentiable and convex. A motivation can come from Bayesian statistics where, given some data $\y\in \Real^{d_y}$, the target is the posterior
\begin{align}\label{eq:targetBayesian}
\widetilde\pi(\x) \propto p(\y|\x) p(\x).
\end{align}
By comparing Eq.~\eqref{eq:targetConvex} and Eq.~\eqref{eq:targetBayesian}, we can choose $p(\y|\x) \propto \exp(-f(\x))$, and 
$p(\x) \propto \exp(-g(\x))$ \cblue{as a log-concave prior distribution}. For instance, $g(\x)$ can encode a sparsity inducing prior $g(\x) = \alpha \|\x\|_1$. Our proposed algorithm can tackle more generic problems where $\pi(\x)$ is decomposed into a product of smooth and nonsmooth parts.

\subsection{Multiple importance sampling}
\label{sec_mis}
Multiple importance sampling (MIS)  is a Monte Carlo method for approximating distributions and integrals by simulating samples from a  set of proposals $\{q_n(\x)\}_{n=1}^N$ \cite{robert1999monte,liu2001monte,mcbook,elvira2019generalized}. A common setup simulates $N$ samples, one from each proposal, i.e., for $n=1,\ldots,N$:
\begin{enumerate}
    \item \textbf{Sampling.} Generate samples $\x_n \sim q_n(\x)$.
    \item \textbf{Weighting.} Two common schemes are:
    \begin{itemize}
        \item {Standard MIS (s-MIS)}: $w_n = \frac{\pi(\x_n)}{q_n(\x_n)}$.
        \item {Deterministic Mixture MIS (DM-MIS)}: $w_n = \frac{\pi(\x_n)}{\psi(\x_n)}$, where $\psi(\x) = \frac{1}{N} \sum_{j=1}^N q_j(\x)$ is the mixture PDF.
    \end{itemize}
\end{enumerate}

Both schemes allow the construction of the classical IS estimators: (a) the unnormalized IS (UIS) estimator, $\widehat{I} = \frac{1}{NZ} \sum_{n=1}^N w_n h(\x_n)$, which requires the normalization constant $Z$ to be known; and (b) the self-normalized IS (SNIS) estimator, $\widetilde{I} = \sum_{n=1}^N \bar{w}_n h(\x_n)$, where $\bar{w}_n = w_n / \sum_{j=1}^N w_j$.
However, the UIS estimator with DM-MIS always outperforms s-MIS in terms of variance reduction as shown in \cite{elvira2019generalized}. Different efficient \cblue{weighting} schemes that also reduce variance have been proposed in \cite{elvira2015efficient,elvira2016heretical,elvira2016overlapping}. 

\subsection{Adaptive importance sampling}
Adaptive importance sampling (AIS) iteratively improves the proposal distributions to reduce estimator variance (see a review in \cite{bugallo2017adaptive}). It adds a third step to MIS, after sampling and weighting in which, typically, the parameters of a mixture proposal are adapted. A key family of AIS algorithms is \emph{population Monte Carlo} (PMC), where this third adaptation step updates the location parameters of the proposals by performing resampling from the current weighted particles. The standard PMC algorithm is described in \cite{cappe2004population}, and further extensions are provided in \cite{cappe2008adaptive,elvira2017improving,elvira2022optimized}.

\subsection{The proximal operator and proximal methods}
 
Proximal methods are a powerful set of techniques that can be used to optimize general cost functions, possibly involving nonsmooth terms \cite{combettes2011proximal, parikh2014proximal}. These algorithms utilize \textit{proximity operators} in order to move towards a fixed-point solution. The proximity operator of a function $g \in \Gamma_0(\bR^{d_x})$ (i.e., the set of proper, lower semicontinuous, convex functions from $\bR^{d_x}$ to $\bR \cup \{+\infty\}$), at a point $x \in \bR^{d_x}$, is defined as
\begin{equation}
\prox_g(\x) = \argmin_{\z\in\bR^{d_x}} g(\z) + \frac{1}{2} \|\z - \x\|_2^2.
\label{eq:prox}
\end{equation}

Our proposed adaptation scheme builds upon the class of proximal gradient methods, whose core principles are now reviewed. 

\subsubsection{The proximal gradient algorithm}
Consider the minimization of $f + g$ over $\bR^{d_x}$, where $f$ is differentiable, and $g \in \Gamma_0(\bR^{d_x})$. 
Given some $\x^{(0)} \in \bR^{d_x}$, the proximal gradient algorithm iterates as
\begin{equation}
(\forall t \in \mathbb{N}) \quad \x^{(t)} = \prox_{\gamma g}\left(\x^{(t-1)} - \gamma \nabla f(\x^{(t-1)})\right), \label{eq:pgrad}
\end{equation}
where $\gamma>0$ is a step-size of the algorithm
set to obtain convergence of the sequence \eqref{eq:pgrad} to a solution to the problem. For instance, if $f \in \Gamma_0(\bR^{d_x})$ and its gradient is $L$-Lipschitz continuous, $(\x^{(t)})_{t \in \mathbb{N}}$ converges to a minimizer of $f+g$, for $\gamma \in (0,2/L)$ \cite{chen1997convergence,tseng2000modified}. 

\subsubsection{Scaled forms for proximal gradient method}
As a first-order method, the algorithm \eqref{eq:pgrad} can display slow convergence. 
Various accelerated forms of proximal gradient algorithm have been investigated. In our adaptation strategy, we will rely on scaled proximal gradient \cite{combettes2014variable,bonnans1995family}, modifying the underlying metric in \eqref{eq:prox}, as follows. Let $\bA \in \bR^{d_x \times d_x}$, symmetric definite positive (SDP). The proximal operator computed at $\x \in \bR^{d_x}$, of $g \in \Gamma_0(\bR^{d_x})$, relative to the metric induced by $\bA$, is
\begin{equation}
\prox_{\bA,g}(\x) = \argmin_{\z\in\bR^{d_x}} g(\z) + \frac{1}{2} (\z - \x)^\top \bA (\z - \x).
\label{eq:proxmetric}
\end{equation}
This new definition yields the following algorithm (again with $x^{(0)} \in \bR^{d_x})$, whose convergence has been explored for instance in \cite{BeckerFadili,Bonettini}, for various classes of $f$ and $(\bA^{(t)})_{t \in \mathbb{N}}$:
\begin{equation}
(\forall t \in \mathbb{N}) \quad \x^{(t)} = \prox_{\bA^{(t)},g}\left(\x^{(t-1)} - \bA^{(t)} \nabla f(\x^{(t-1)})\right).
\label{eq:algoproxmetric}
\end{equation}
\subsubsection{Extension to non-convex settings}
The  algorithms above have been extended this last decade, to the non-convex settings, under specific assumptions on $(f,g)$ to guarantee the well-posedness of the iterates, in particular, to ensure that the proximal map of $g$ is well-defined (uniqueness/existence). Convergence guarantees to critical points, can be obtained under the \cblue{Kurdyka-\L{}ojasewicz} framework~\cite{attouch2009convergence,chouzenoux:hal-00789970,attouch2013convergence}.

\section{Proximal Newton adaptive importance sampler}
\label{sec_contribution}

\begin{table}[!t]
  \centering
  {
  \caption{{{\color{black} \acro \ algorithm.}}}
  {
  \begin{tabular}{|p{0.95\columnwidth}|}
    \hline
    \footnotesize
    \begin{enumerate}
      \item {\bf [Initialization]}: Set $\sigma>0$, $(N,K,T) \in \mathbb{N}^+$, $\{\bnu_n\}_{n=1}^N$. For $n=1,\ldots,N$, select the initial adaptive parameters ${\bm \mu}_n^{(1)} \in \mathbb{R}^{d_x}$ and $\bSigma_n^{(1)} = \sigma^2 \mathbf{I}_{d_x}$. 
      \vspace*{6pt}
      \item {\bf[For $\bm t \bm= \bm 1$ to  $\bm T$]}: 
      \begin{enumerate}
        \item {\bf [Sampling]}: Simulate $NK$ samples as
          \begin{equation} 
            \x_{n,k}^{(t)} \sim q_n^{(t)}(\x;\bmu_n^{(t)},\bSigma_n^{(t)},\bnu_n)
            \label{eq_drawing_part_pmc}
          \end{equation}
          
          with $n=1,\ldots,N$, and $k=1,\ldots,K$. 
          \item {\bf [Weighting]}: Calculate the normalized IS weights as
          \begin{equation} 
              w_{n,k}^{(t)} = \frac{\pi(\x_{n,k}^{(t)})}{\frac{1}{N}\sum_{i=1}^N  q_i^{(t)}(\x_{n,k}^{(t)})}.
          \label{is_part_weights}
          \end{equation}
          \item {\bf [Adaptation]}: Adapt the location and scale parameters of the proposal
          \begin{enumerate}
              \item  {\bf [Resampling step]} {Resample $N$ proposals densities from the pool of $NK$ weighted samples at the iteration $t$.  The means and scales of the resampled proposals are denoted as $\widetilde {\bm \mu}_n^{(t)}$ and $\widetilde{\bSigma}_n^{(t)}$, respectively. See Section \ref{sec_resampling} for explicit definitions of the notation.}  
        \item {\bf [Optimization step]} Adapt the proposal parameters $\{({\bm \mu}_n^{(t+1)},\bSigma_n^{(t+1)})\}_{n=1}^N$ according to \eqref{eq_mean_adapt}-\eqref{eq_cov_adapt}.
            \end{enumerate} 
\end{enumerate}
    \item {\bf [Output, $\bm t \bm =\bm T$]}: 
        Return the pairs $\{\x_{n,k}^{(t)}, {w}_{n,k}^{(t)}\}$, for
        $n=1,\ldots,N$, $k=1,\ldots,K$ and $t=1,\ldots,T$.
    \end{enumerate} \\
    \hline 
\end{tabular}\label{PMC_framework_new}
}
}
\end{table}

We display our proposed \acro\; algorithm in Table \ref{PMC_framework_new} to adapt the set of $N$ proposals for $t=1,...,T$ iterations. We denote them as  $\{ q_t(\x;\bmu_n^{(t)},\bSigma_n^{(t)}, \bnu_n) \}_{n=1}^N$, where $\bmu_n^{(t)}$ and  $\bSigma_n^{(t)}$ are the adapted location and scale parameters, respectively, and $\bnu_n$ are the static parameters (e.g., degrees of freedom in Student's t-distributions). The algorithm runs over $T$ iterations to finally produce a set of $KNT$ weighted samples, enabling the construction of IS estimators. Three steps are performed at each iteration $t$, namely sampling $K$ samples per proposal (Step 2a), weighting the samples with the DM-MIS scheme (Step 2b), and adapting the location and scale parameters (Step 2c). The last step involves a resampling procedure, the mean adaptation, and the covariance adaptation; each of these mechanisms is detailed in the next three subsections.

\subsection{Resampling procedure}
\label{sec_resampling}
As in most PMC algorithms, the resampling step simulates the location of the  $N$ next proposals. The global resampling (GR) creates a pool of $NK$ samples, with weights proportional to \eqref{is_part_weights} and normalized over the $NK$ samples; GR resamples $N$ location parameters with replacement from the pool. The local resampling (LR) performs instead exactly one drawing from $N$ different pools, each of them composed of the $K$ samples simulated from each proposal, and with weights proportional to \eqref{is_part_weights} and normalized over the $K$ samples. GR generally converges faster but loses diversity, while LR keeps diversity (exactly one sample per proposal survives) at the cost of keeping proposals that may not cover significant probability mass of the target (see more details in~\cite{elvira2017improving,elvira2017population}). Here, we follow a hybrid approach, called  \emph{``glocal''} resampling (GLR)~\cite{elvira2022optimized}, which is well tailored to be followed by the optimization step described in the next section. 
In GLR, most iterations perform an LR step except that every $\Delta \in \cblue{\mathbb{N}^+}$ iterations, a GR step is performed instead.

\subsection{Mean and covariance adaptation}

\cblue{The mean adaptation follows the optimization strategy in~\eqref{eq:algoproxmetric}. Specifically, at each iteration $t$,}  we compute the proposal mean ${\bmu}_n^{(t+1)}$, for every $n \in \{1,\ldots,N\}$ by performing one step of the proximal Newton algorithm from~\cite{BeckerFadili} initialized at $\tilde{\bmu}_n^{(t)}$. The adapted mean, for a given $(n,t)$, is 
\begin{equation}
{\bm \mu}_n^{(t+1)}  = \prox_{\bA({\widetilde{\bmu}}_n^{(t)})^{-1},  g} \left( \widetilde {\bm \mu}_n^{(t)} \ -   \bA({\widetilde{\bmu}}_n^{(t)}) \nabla f( \widetilde {\bm \mu}_n^{(t)} ) \right). \label{eq_mean_adapt}
\end{equation}
Hereabove, $\bA(\widetilde{\bmu}_n^{(t)})$ is an SDP matrix of $\mathbb{R}^{d_x \times d_x}$ that is scaling the proximal gradient update. Following~\cite{BeckerFadili}, we define
\begin{equation}
\label{eq:covadapt1a}
\bA(\widetilde{\bmu}_n^{(t)}) = \theta_n^{(t)}  \mathbf{\Gamma}(\widetilde{\bmu}_n^{(t)}),
\end{equation}
with the Newton-like matrix
\begin{equation}
\label{eq:covadapt1b}
\mathbf{\Gamma}(\widetilde{\bmu}_n^{(t)}) = 
\begin{cases}
\left(\nabla^2  f(\widetilde{\bmu}_n^{(t)})\right)^{-1}, & \text{if} \; \nabla^2 f(\widetilde{\bmu}_n^{(t)})\succ 0,\\
{\widetilde \bSigma_{n}^{(t)}}, & \text{otherwise}.
\end{cases}
\end{equation}
Since $f$ might be non convex (i.e., smooth part of the target is not log-concave), we introduce two safe rules. First, if the Hessian of $f$ at $\tilde{\bmu}_n^{(t)}$ is non invertible, \cblue{we instead set the scaling matrix to the} covariance of the proposal that generated the sample. Second, we introduced in \eqref{eq:covadapt1a} a damped factor $\theta_n^{(t)}\in (0,1]$, tuned according to a backtracking scheme \cite{Bonettini} to avoid the degeneracy of the proximal Newton iteration, and thus of our adaptation scheme. Initialized with unit stepsize value, we apply a reduction factor $\tau = 1/2$ until the target increases.

\cblue{In order to be consistent with the mean adaptation, the covariance matrix of the proposal is also adapted as}
\begin{equation}
\bSigma_{n}^{(t+1)} = \bA(\widetilde{\bmu}_n^{(t)}). \label{eq_cov_adapt}
\end{equation}
 
We emphasize that our proposed proposal adaptation method, in \eqref{eq_mean_adapt}-\eqref{eq_cov_adapt}, is reminiscent from the preconditioned proximal unadjusted Langevin algorithm (PP-ULA) introduced in~\cite{corbineau:hal-02474585} in the context of an MCMC sampler for ultrasound imaging. In particular, as the authors of \cite{corbineau:hal-02474585} show in their appendix, the considered adaptation can be viewed as the Euler discretization of the Langevin diffusion equation applied to the target $\pi$ with
preconditioning matrix \eqref{eq_cov_adapt}. 
\cblue{The convergence of the scheme \eqref{eq_mean_adapt}, without resampling (i.e., $\tilde{\bmu}_n^{(t)} \equiv \bmu_n^{(t)}$), to a minimizer of $f+g$ has been established in the convex setting in \cite{LeeProxNewton}. The non-convex case has been studied, for instance in \cite{chouzenoux:hal-00789970,repetti2021variable}, under extra assumptions on the scaling matrix and stepsize, later extended to the stochastic setting in \cite{Fort2023}.}

\subsection{Practical calculation of the proximal step}
In most useful applications of PNAIS, the scaling metric given by \eqref{eq:covadapt1a}-\eqref{eq:covadapt1b} might be non trivial (e.g., not diagonal) and/or function $g$ might be complicated (e.g., non separable), hence the evaluation of the proximity operator in \eqref{eq_mean_adapt} might not take a closed-form, and an inner solver is necessary. \cblue{Several approaches are possible.} Here, we opted for the dual forward-backward algorithm \cite{CombettesDFB}, provided in the supplementary material. This algorithm only requires the expression for the proximity operator of $g$, which is simple for a wide class of examples.\footnote{See https://proximity-operator.net/} Note that parallelized/distributed implementations for the dual forward backward algorithm are also available \cite{abboud:hal-03684063}.

\section{Numerical examples}
\label{sec_numerical}
We now present two numerical examples concerning a distribution constrained in the simplex (Example A) and a non-differentiable target (Example B), both with $d_x = 2$. \cblue{We refer the reader to our supplementary file, for an Example C, on a high-dimensional setting.}
All our results are averaged over 100 
independent runs. {We use Gaussian proposals, with the location parameters initialized with random elements from $[0,1]^{d_x}$.} 
Except otherwise stated, we set $\sigma=1$ (i.e., initial standard deviation of isotropic covariances). 
In particular, the initial samples do not necessarily belong to the domain of $g$. The GLR of Section \ref{sec_resampling} has a parameter $\Delta=5$, and we set $(N,K,T) = (50,20,20)$. {We refer the reader to our supplementary material, for numerical results using the LR strategy.} 

\noindent\textbf{Ablation study.} Our experimental analysis takes the form of an ablation study. Namely, we compare the performance of PNAIS, to modified versions of it, where some features have been changed/discarded. Our competitors are as follows:
\begin{itemize}
    \item DM-PMC: We discard step c) of PNAIS algorithm, that is we do not perform any proposal adaptation. 
    \item PNAIS-nocov: We simplify \eqref{eq_cov_adapt}, and set, for every $(n,t)$, $\bSigma_n^{(t)} = \sigma^2 \mathbf{I}_{d_x}$.
    \item PNAIS-rcov: Instead of \eqref{eq_cov_adapt}, we use the robust covariance adaptation from \cite{Laham2018}.
    \item PNAIS-grad: We modify the mean adaptation \eqref{eq_mean_adapt} using the standard proximal gradient step (i.e., no Newton scaling is performed), 
    \begin{align}
{\bm \mu}_n^{(t+1)}  &= \prox_{\theta_n^{(t)}  g} \left( \widetilde {\bm \mu}_n^{(t)} \ -   \theta_n^{(t)} \nabla f( \widetilde {\bm \mu}_n^{(t)} ) \right),
\end{align}
with stepsize $ \theta_n^{(t)}$ computed following a similar backtracking procedure than PNAIS.
\end{itemize}

\subsection{Example A: Gaussian mixture over the simplex}
 
\begin{figure}
    \centering
    \begin{tabular}{c@{}c}
    \includegraphics[height=4cm]{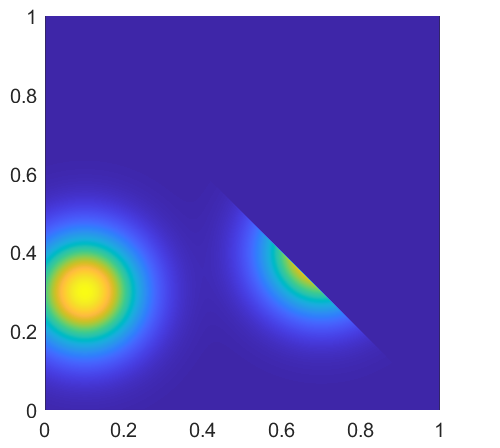}
    &
        \includegraphics[height=4.15cm]{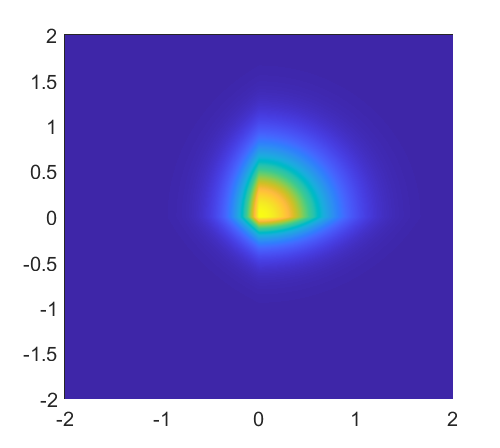}
    \end{tabular}
        \caption{(left) Target function for Example A. The Target equals 0 outside the unit simplex set. (right) Target function for Example B.}
    \label{fig:ExAB}
\end{figure}

We consider a truncated version of an equally weighted mixture of bivariate Gaussian distributions, with means $[0.1,0.3]^\top$ and $[0.7,0.4]^\top$ and  both covariances equal to $[0.01,0 ; 0,0.01]$. The mixture is truncated to be defined only in the unit simplex i.e., $\x \geq 0$, $x_1 + x_2 \leq 1$, as displayed in Fig. \ref{fig:ExAB} (left). We define $f$ as the neg-logarithm of the Gaussian mixture distribution, and $g$ as the indicator function of the simplex set. The proximity operator of $g$ is the projection over this set. The ground truth values, for the mean, second-order moment, and normalization constants, determined by numerical integration with a rough grid, are, respectively, $E_{\tilde{\pi}}[X] = [0.2369, 0.3023]^\top$, $E_{\tilde{\pi}}[X^2] = [0.1024, 0.1005]^\top$ and $Z = 0.5398$. The results for estimating these values, are summarized in Table \ref{table_exp1} (left), in terms of relative mean squared error (MSE). The proposed PNAIS clearly outperforms its competitors. In particular, DM-PMC is largely behind in terms of performance. Moreover, the ablated versions of PNAIS present limitations.

\begin{table}[h!]
\scriptsize
    \centering
    \scalebox{0.9}{
    \begin{tabular}{|m{0.1cm}|m{0.6cm}|m{1cm}|m{1cm}|m{1cm}||m{1cm}|m{1cm}|m{1cm}|}
    \cline{3-8}
\multicolumn{2}{c }{} & \multicolumn{3}{|c||}{Example A} & \multicolumn{3}{|c|}{Example B} \\
    \cline{3-8} 
   \multicolumn{2}{c|}{} & $E_{\tilde{\pi}}[X]$ & $E_{\tilde{\pi}}[X^2]$ & $Z$ & $E_{\tilde{\pi}}[X]$ & $E_{\tilde{\pi}}[X^2]$ & $Z$  \\
   \hline
   \multirow{3}{*}{\begin{sideways}\tiny{DM-PMC}\end{sideways}} & $\sigma$ = $1$& \num{1.1229e-04} & \num{6.0219e-5}& \num{1.7961e-3} & \num{2.9575e-5}& \num{1.7897e-5} & \num{1.0896e-5} \\
   \cline{2-8}
 & $\sigma$ = $3$ & \num{2.1618e-03} & \num{1.0725e-3}& \num{3.5759e-2} & \num{2.747e-4}& \num{1.2272e-4} & \num{1.1566e-4} \\
   \cline{2-8}
 & $\sigma$ = $5$ & \num{2.9550e-01} & \num{6.6993e3} & \num{1.7467e-1}  & \num{7.50e-4}& \num{4.0933e-4}& \num{3.6361e-4}\\
   \hline
      \multirow{3}{*}{\begin{sideways}\tiny{PNAIS-grad}\end{sideways}} & nocov & \num{1.3948e-4}& \num{7.3103e-5}& \num{1.4003e-3}  & \num{2.6764e-5}& \num{1.5602e-5}& \num{7.2209e-6} \\
   \cline{2-8}
 & rcov & \num{1.9595e-5} & \num{7.3205e-6} &\num{1.1227e-4} & \num{6.6771e-5}& \num{7.4483e-5} &\num{3.7529e-6} \\
   \cline{2-8}
 & \eqref{eq_cov_adapt} & \num{5.6850e-3}& \num{3.649e-3} & \num{8.8743e-2} & \textbf{\num{1.0975e-5}}&\textbf{\num{1.1264e-5}} & \num{1.067e-6} \\
   \hline
         \multirow{3}{*}{\begin{sideways}\tiny{PNAIS}\end{sideways}} & nocov & \num{1.2979e-4}& \num{5.8311e-5} & \num{1.1817e-3} & \num{3.7755e-5}& \num{2.2768e-5}& \num{1.0551e-5} \\
   \cline{2-8}
 & rcov & \num{8.9782e-6}& \num{4.3682e-6}& \num{4.768e-5}& \num{2.3211e-3}& \num{2.0778e-3}& \num{4.6299e-4} \\
   \cline{2-8}
 & \eqref{eq_cov_adapt} & \textbf{\num{5.018e-6}}& \textbf{\num{2.4524e-6}} & \textbf{\num{1.627e-5}}& \num{1.5633e-5}& \num{1.8061e-5}&  \textbf{\num{5.6357e-7}}\\
   \hline
    \end{tabular}
    }
    \caption{Relative MSE for Examples A (left) and B (right).}
    \label{table_exp1}
\end{table}

\subsection{Example B: Gaussian likelihood with sparse prior}
{We consider a target of the form \eqref{eq:targetBayesian}, with $p(\y|\x)$ a Gaussian distribution with mean $[0.5,0.5]^\top$ and covariance $[0.25,0 ; 0,0.25]$, and $p(\x) \propto \exp(-\alpha \| \x\|_1)$, with $\alpha=2$.} 
The target function is displayed in Figure \ref{fig:ExAB} (right). We define $f$ as the neg-logarithm of the Gaussian distribution, and $g = \alpha \| \cdot\|_1$. The proximity operator of $g$ is the soft-thresholding operator with scale $\alpha$. We aim at estimating the ground truth integrals $E_{\tilde{\pi}}[X] = [0.2025 ;0.2025]^\top$, $E_{\tilde{\pi}}[X^2] = [0.252,0.252]^\top$, and $Z = 0.1641$, using PNAIS or its variants.  The results are summarized in Table \ref{table_exp1} (right). In this example, the first and \cblue{second} order moments are \cblue{well estimated by all methods. PNAIS shows superior performance with the Hessian-based covariance adaptation and  without metric acceleration in the proximal gradient mean adaptation.} The normalization constant estimation is difficult in this example, due to an \cblue{asymmetric} target shape with highly non Gaussian shape. Here, the proposed PNAIS is largely superior to its competitors, showing the interest of both adaptation schemes to accurately explore the target. 
\cblue{In both examples, we noted a similar time complexity of PNAIS, when compared to its competitors.}

 \section{Conclusion}
 \label{sec_conclusion}
 
 The success of AIS heavily depends on the effective adaptation of proposal distributions. In this paper, we proposed a proximal Newton adaptive importance sampler, an algorithm that exploits geometric information for estimating expectations with respect to non-smooth target distributions. By leveraging a scaled Newton proximal gradient, the algorithm adapts multiple proposals to approximate targets that are partially non-differentiable. We have shown its good performance in two challenging numerical experiments. 
\cblue{Future work will explore reduced complexity extensions of PNAIS in higher dimensions, e.g., for inference in Bayesian neural networks~\cite{huang:hal-04210696}.}

\bibliographystyle{abbrv}

\onecolumn
\newpage

\section*{Suplementary material}

\subsection*{Dual forward-backward algorithm to compute the proximal step}

We summarize in Table S-I the iterations of the dual forward-backward algorithm [15], to compute the proximity operator of a function $g \in \Gamma_0(\mathbb{R}^{d_x})$, within the metric induced by an SDP matrix $\bA^{-1}$, at some point $\tilde{\bxi} \in \mathbb{R}^{d_x}$. Let $\| \bL \|$ denote the spectral norm of matrix $\bL$. Note that we use a particular form of the method in [16], with a single proximable term, a unit stepsize, and a specific choice for the initialization. The sequence $(\bL \bxi_j)_{j\geq 1}$ converges to the solution of the sought problem (as a consequence of [15, Theorem 2.2], and continuity of the linear operator $\bL$). In practice, a few iterations are needed to reach stability, and one can exit the loop as soon as the relative norm difference between two consecutive iterates of $(\bxi_j)_{j\geq 1}$ gets lower than some low value (typically, $10^{-7}$).

\begin{table}[h]
\renewcommand\thetable{S-I}
  \centering
  {
  \caption{{{\color{black} Dual Forward Backward algorithm to compute $\text{prox}_{\bA^{-1},g}(\tilde{\bxi})$.}}}
  {
  \begin{tabular}{|p{0.85\columnwidth}|}
    \hline
    \footnotesize    {
    \begin{enumerate}
\item \textbf{[Initialization]}: Set $\bA \in \bR^{d_x \times d_x}$ SDP, $\tilde{\bxi} \in \bR^{d_x}$, and $g \in \Gamma_0(\bR^{d_x})$. Set iteration number $J>0$. 
\item[] Set $\bL = \bA^{1/2}$, $\tilde{\bzeta} = \bL^{-1} \tilde{\bxi}$, $\rho = \| \bL\|^2$, $\bzeta_1 = \bL \tilde{\bxi}$.
\item \textbf{[For $j=1,\ldots,J$]}:
\begin{enumerate}
    \item[] $\bxi_j = \tilde{\bzeta} - \bL^\top \bzeta_j$ 
    \item[] $\tilde{\bzeta}_j = \bzeta_j + \rho^{-1} \bL \bxi_j$
    \item[] $\bzeta_{j+1} = \tilde{\bzeta}_j - \rho^{-1} \text{prox}_{\rho g}(\rho \tilde{\bzeta}_j)$
\end{enumerate}
\item \textbf{[Output]}: Return $\bL \bxi_J$
    \end{enumerate}}\\
    \hline
    \end{tabular}
    }
    }
    \label{tab:dfb}
    \end{table}
    \vspace*{-0.5cm}

\subsection*{Additional results}
\cblue{\subsubsection*{Local resampling strategy}}

In Tables \ref{table_exp1lr} and \ref{table_exp2lr}, we provide the results for Examples A and B, respectively, using LR resampling strategy, instead of the GLR one, in step c)i) of PNAIS algorithm. Again, the proposed method displays excellent performance on both examples. In Example A, it reaches the best results among the competitors. In Example B, the first-order version of PNAIS is slightly better (i.e., lower MSE) for the three estimated quantities. Let us now compare the results to those obtained with the GLR strategy, on the same tasks by inspecting Table II, in the main file. In Example A, the LR strategy is slightly superior to GLR, in terms of relative MSE, in most cases. In contrast, on Example B, higher performance was obtained when using the GLR strategy, in particular for estimating the mean and the normalization constant of the target. 

\begin{table} [h!]
\renewcommand\thetable{S-II}
\scriptsize{
\setlength{\tabcolsep}{2pt}
\def\marginwidth{1.5mm}
\begin{center}
\begin{tabular}{|c||c|c|c||c|c|c||c|c|c||}                                 
\hline 
&  \multicolumn{3}{c }{DM-PMC} &  \multicolumn{3}{|c|}{PNAIS-grad}  & \multicolumn{3}{|c||}{PNAIS}  \\
\cline{2-10}
&  $\sigma=1$  & $\sigma=3$   & $\sigma=5$  &  nocov  & rcov & (13)  & nocov  & rcov  &  (13) \\
\hline
\hline
$E_{\tilde{\pi}}[X]$ &  3.3956e-04 &1.1882e-02 &1.5891e+00  &  2.1561e-04   & 2.1532e-05 &  3.7829e-03 &  1.0816e-04  & 1.4407e-05  & \textbf{4.2893e-06}  \\
\hline 
$E_{\tilde{\pi}}[X^2]$ & 1.5761e-04 &5.2274e-03 &1.9780e+04  &   1.0776e-04  & 8.9325e-06 & 2.1195e-03  &  4.7324e-05  & 6.6360e-06  &   \textbf{2.1160e-06}\\
\hline 
Z &  1.1815e-03 &1.1186e-01 &5.5812e-01 &  2.2864e-03  & 2.0005e-04  &  1.3271e-01 &  1.9897e-03 & 5.6512e-05  & \textbf{1.0255e-05}\\ 
\hline
\end{tabular}
\end{center}
\caption{\textbf{Example A.} Relative MSE using LR approach. 
\label{table_exp1lr}
}
}
\end{table}


\begin{table} [h!]
\renewcommand\thetable{S-III}
\scriptsize{
\setlength{\tabcolsep}{2pt}
\def\marginwidth{1.5mm}
\begin{center}
\begin{tabular}{|c||c|c|c||c|c|c||c|c|c||}                                 
\hline 
&  \multicolumn{3}{c }{DM-PMC} &  \multicolumn{3}{|c|}{PNAIS-grad}  & \multicolumn{3}{|c||}{PNAIS}  \\
\cline{2-10}
&  $\sigma=1$  & $\sigma=3$   & $\sigma=5$  &  nocov  & rcov & (13)   & nocov  & rcov  &  (13)  \\
\hline
\hline
$E_{\tilde{\pi}}[X]$ & 2.6789e-05 &2.2452e-04 &8.8604e-04 &  3.6815e-05  & 5.4597e-03  & \textbf{1.4897e-05} &  2.4109e-05  & 7.0972e-03 & 2.3900e-05 \\
\hline
$E_{\tilde{\pi}}[X^2]$ & 1.9983e-05 &1.6666e-04 &4.3758e-04  & 2.0217e-05 & 1.1799e-02  & \textbf{1.0674e-05}  &  1.7564e-05   &  1.1835e-02 & 2.8333e-05   \\
\hline
Z & 7.5993e-06 &1.8578e-04 &3.9083e-04  & 8.2244e-06  &  5.3964e-04 & \textbf{1.2680e-06} &  5.2137e-06  & 1.0151e-03& {2.2691e-06} \\
\hline
\end{tabular}
\end{center}
\caption{\textbf{Example B.} Relative MSE using LR approach. 
\label{table_exp2lr}
}
}
\end{table}
 
\cblue{\subsubsection{High-dimensional truncated banana-shaped distribution}}

\cblue{In Table \ref{table_exp2ls},  we consider the problem of estimating the mean of a truncated banana-shaped distribution in up to $d_x=100$ dimensions, constrained to the domain $\|x\|_2 \leq 4$ (see Fig. S-V(left)). The target distribution is defined as in [26, Sec. V-B]. The projection onto the $\ell_2$-ball constraint has a closed form (see for instance, https://proximity-operator.net/). The same settings as in Examples A and B, were used, for $(N,K,T, \sigma)$, and all methods implement GLR resampling. The Table S-IV displays the results of the different methods, in terms of relative MSE, for the estimation of the mean of the truncated distribution, in the cases $d_x \in \{2,10, 50\}$ ($\times$ means that the method does not manage to stabilize). Note that the first coordinate of the mean is discarded in the MSE computation, as the ground truth is not known for this example. The table shows the clear superiority of our approach. In Fig. S-V(right), we show the computational time when running the methods all implemented with Matlab R2023a code, on a Intel(R) Core(TM) i7-8650U CPU @ 1.90GHz 2.11 GHz, 16Go RAM, for different problem sizes, $d_x \in \{2,5,10,20,50,100\}$. We observe that the complexity increases with dimension in all methods. Our method does not significantly increase the computation time compared to competitors, probably because the complexity is dominated by the sampling and resampling steps. 
}

\begin{table} [h!]
\renewcommand\thetable{S-IV}
\scriptsize{
\setlength{\tabcolsep}{2pt}
\def\marginwidth{1.5mm}
\begin{center}
\cblue{
\begin{tabular}{|c||c|c|c||c|c|c||c|c|c||}    \cline{2-10}
\multicolumn{1}{c }{ } & \multicolumn{3}{|c }{DM-PMC} &  \multicolumn{3}{|c|}{PNAIS-grad}  & \multicolumn{3}{|c||}{PNAIS}  \\
\cline{2-10}
 \multicolumn{1}{c|}{ } & $\sigma=1$  & $\sigma=3$   & $\sigma=5$  &  nocov   & rcov & (13)   & nocov  & rcov &  (13)  \\
 \hline
$d_x=2$ & 9.9112e-05 &3.5901e-05 &3.8034e-04 &  1.4155e-04  & 3.4049e-05 & 9.4337e-03 &  1.4751e-04 &  7.1918e-05 & \textbf{6.7664e-06} \\ 
\hline
$d_x = 10$ & 1.7353e-03 &5.6920e+00 &1.3211e+01 &  4.0847e+00  & 4.3783e-02  & 6.9631e-02 & 7.8396e-01 & 3.7315e-02 & \textbf{1.0493e-04} \\ 
\hline
$d_x = 50$ & 8.7256e-01 &4.7679e+00 &1.5457e+01 &  3.4186e+00  & $\times$  & 3.0231e+00 & 2.0223e+00 & $\times$ & \textbf{4.0775e-01} \\ 
\hline
\end{tabular}
}
\end{center}
\caption{\textbf{Example C.} Relative MSE on the mean over all coordinates but the first (for which no ground truth is available). 
\label{table_exp2ls}
}
}
\end{table}

\begin{figure}[h]
\renewcommand\thefigure{S-V}
    \centering  
    \begin{tabular}{cc}
        \includegraphics[height=6cm]{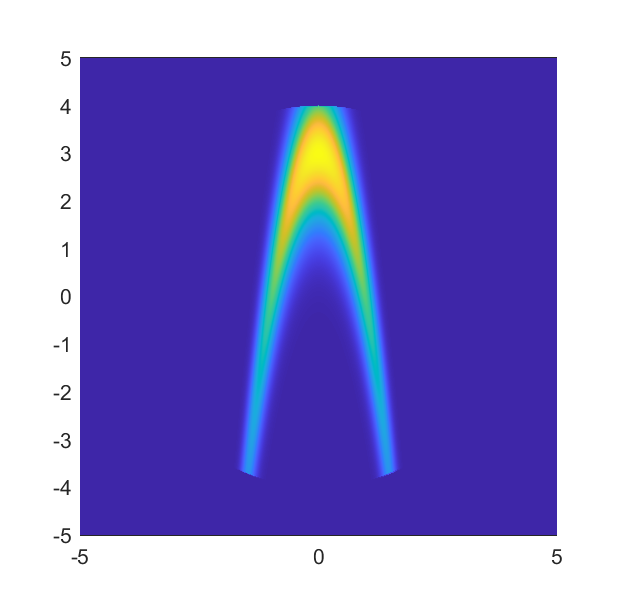} & \includegraphics[height=6cm]{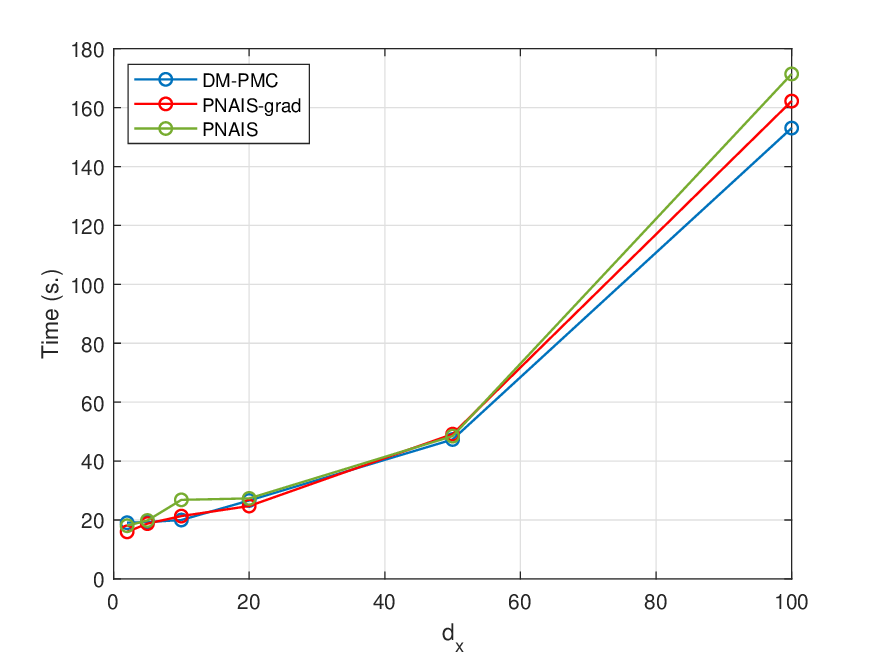} 
    \end{tabular}
\caption{\cblue{\textbf{Example C.} Left: two first dimensions of the target distribution with $\ell_2$ ball constraint (with radius 4). Right:  Computational time per run as a function of $d_x$ for DM-PMC ($\sigma=1$), PNAIS-grad (covariance adaptation as in (13)), and the proposed PNAIS method.}}
    \label{fig:enter-label}
\end{figure}

\end{document}